\documentclass[pra,aps,nofootinbib,notitlepage,twocolumn]{revtex4-1}

\usepackage{amsthm}
\usepackage{amsmath,bm}
\usepackage{amssymb}
\usepackage{amsfonts}
\usepackage{graphicx}

\usepackage{txfonts}

\usepackage[colorlinks=true,linkcolor=blue,citecolor=blue,urlcolor=blue]{hyperref}

\begin{document}

\title{Superresolution Limits from Measurement Crosstalk}
\author{Manuel Gessner}
\affiliation{Laboratoire Kastler Brossel, ENS-Universit\'{e} PSL, CNRS, Sorbonne Universit\'{e}, Coll\`{e}ge de France, 24 Rue Lhomond, 75005, Paris, France}
\author{Claude Fabre}
\affiliation{Laboratoire Kastler Brossel, ENS-Universit\'{e} PSL, CNRS, Sorbonne Universit\'{e}, Coll\`{e}ge de France, 24 Rue Lhomond, 75005, Paris, France}
\author{Nicolas Treps}
\affiliation{Laboratoire Kastler Brossel, ENS-Universit\'{e} PSL, CNRS, Sorbonne Universit\'{e}, Coll\`{e}ge de France, 24 Rue Lhomond, 75005, Paris, France}
\date{\today}

\begin{abstract}
Superresolution techniques based on intensity measurements after a spatial mode decomposition can overcome the precision of diffraction-limited direct imaging. However, realistic measurement devices always introduce finite crosstalk in any such mode decomposition. Here, we show that any nonzero crosstalk leads to a breakdown of superresolution when the number $N$ of detected photons is large. Combining statistical and analytical tools, we obtain the scaling of the precision limits for weak, generic crosstalk from a device-independent model as a function of the crosstalk probability and $N$. The scaling of the smallest distance that can be distinguished from noise changes from $N^{-1/2}$ for an ideal measurement to $N^{-1/4}$ in the presence of crosstalk.
\end{abstract}

\maketitle

The precision of optical imaging devices determines the state of the art in microscopy and astronomy. While diffraction affects our ability to resolve small structures and separations by spatially distributed intensity measurements, it does not pose a fundamental limitation. Historical resolution limits of Abbe, Rayleigh, and others address the effect of diffraction but they become irrelevant if the signal-to-noise ratio is high enough~\cite{Goodman}. Moreover, a variety of superresolution techniques have been developed to overcome these limits, e.g., by stimulated-emission microscopy~\cite{Hell}, by making use of homodyne measurements~\cite{Kolobov,DelaubertPRA,DelaubertOptLett}, or by intensity measurements in a transformed basis of modes~\cite{HelstromResolution,TsangReview}.

A systematic approach to the estimation of the spatial separation between two light sources is provided by the formalism of quantum metrology~\cite{Helstrom,Holevo,Paris,Giovannetti}. Tools from quantum information theory allow us to optimize over all conceivable measurement techniques, in order to identify the ultimate quantum limits on precision~\cite{BC94}. For example, it was shown in Ref.~\cite{Tsang} that spatial demultiplexing in transverse electromagnetic (TEM) Hermite-Gauss modes realizes a quantum-optimal measurement for the estimation of the separation between two incoherent sources. These results have been extended to two-dimensional images~\cite{TsangPRA}, thermal states~\cite{NairTsangPRL,LupoPRL}, and more general scenarios~\cite{Rehacek,Backlund,Yu,Napoli,LupoPRL2020}. They have further been implemented experimentally by using interferometric phase-sensitive measurements~\cite{Ling,Steinberg,Parniak}, digital holography~\cite{RehacekDigitalHolography}, or by using a local oscillator in an excited TEM$_{01}$ mode~\cite{Yang,Silberhorn}. A decomposition of the detected light in TEM$_{nm}$ modes with $0\leq n,m\leq 2$ was recently realized~\cite{Pauline} using a multiplane-light-conversion technique~\cite{MPLC}. However, any experimental mode decomposition suffers from unavoidable crosstalk between the modes. So far, theoretical treatments of deviations from the ideal scenario have been limited to misaligned centroids~\cite{Tsang} and electronic detection noise~\cite{Banaszek,Lupo2020}. 

In this Letter, we show that nonzero crosstalk between the detector modes before an intensity measurement leads to a breakdown of superresolution at small source separations. To quantify precision, we introduce the minimal resolvable distance $d_{\min}$ at unit signal-to-noise ratio as a function of the total number $N$ of photons. For the separation between two incoherent point sources, measurement crosstalk implies a change in the scaling of $d_{\min}$ when $N\gg 1$: While in the ideal, noiseless case $d_{\min}/w=N^{-1/2}$, in the presence of crosstalk, we obtain $d_{\min}/w=\alpha N^{-1/4}$, where $w$ is the beam width. An analytical model for weak crosstalk predicts that $\alpha\propto\sqrt{|r|}$, where $|r|^2$ is the crosstalk probability. These analytical results agree with the statistical predictions of a random-matrix model for generic crosstalk, indicating that this scaling is device independent. At low photon numbers, or for widely separated probes, we find that crosstalk is not a fundamental limitation and the measurement of higher-excited modes becomes increasingly relevant.

\begin{figure}[tb]
\centering
\includegraphics[width=.49\textwidth]{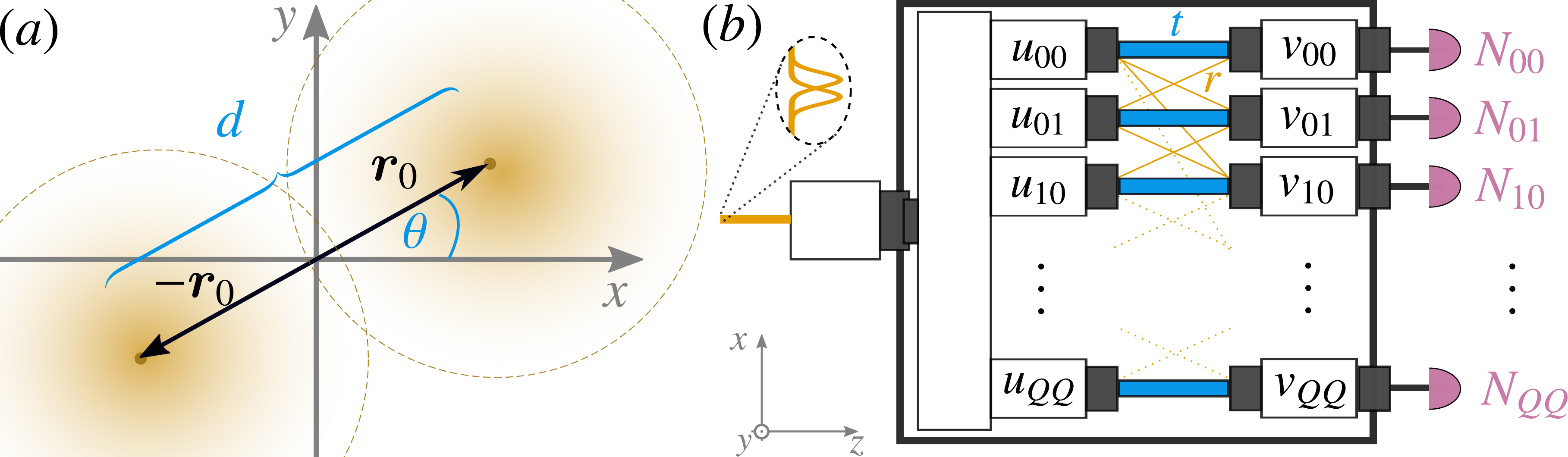}
\caption{(a) Two sources centered at positions $\pm\bm{r}_0$ with distance $d$ create overlapping Gaussian intensity distributions in the image plane. (b) Schematic representation of a mode decomposition with crosstalk. Before an intensity measurement, each mode is either transmitted $(t)$ into the correct output, or reflected $(r)$ into another mode due to crosstalk.}
\label{fig:setup}
\end{figure}

\textit{Mode decomposition and measurement.---}We consider the problem of separating two incoherent point sources of equal intensity, located in a two-dimensional plane at positions $\pm\bm{r}_0$, where $\bm{r}_0=(d/2)(\cos\theta,\sin\theta)$, see Fig.~\ref{fig:setup}~(a). After passing through a diffraction-limited imaging system, the spatial field distribution is, to a good approximation~\cite{Goodman}, described by two overlapping Gaussian profiles $u_{0}(\bm{r}\pm\bm{r}_0)$ with $u_{0}(x,y)=\sqrt{2/(\pi w^2)}\exp(-[x^2+y^2]/w^2)$. The function $u_{0}(\bm{r})$ can be extended to a complete Hermite-Gauss basis $\{u_{k}(\bm{r})\}$ with $k=(n,m)$ and $u_0=u_{00}$. To describe the electromagnetic field in the image plane, it is convenient to introduce two mode bases that are centered at the respective source positions, i.e., $u_{\pm,k}(\bm{r})=u_{k}(\bm{r}\mp\bm{r}_0)$. Introducing corresponding field operators $\{\hat{b}_{\pm,k}\}$, both bases can be used to represent the electric field operator as $\hat{E}^{(+)}(\bm{r})=\sum_{k}u_{+,k}(\bm{r})\hat{b}_{+,k}=\sum_{k}u_{-,k}(\bm{r})\hat{b}_{-,k}$. We assume that the emitted photons follow a Poisson distribution and the electromagnetic field is described by $M$ independent copies of a quantum state $\hat{\rho}(d,\theta)$ containing $\epsilon\ll 1$ photons. We have
\begin{align}\notag
\hat{\rho}(d,\theta)=\frac{1}{2}\left(\int d\alpha P_+(\alpha)|\alpha\rangle_+\langle\alpha|_++\int d\alpha P_-(\alpha)|\alpha\rangle_-\langle\alpha|_-\right),
\end{align}
where $|\alpha\rangle_{\pm}=\exp(\alpha\hat{b}^{\dagger}_{\pm,0}-\alpha^*\hat{b}_{\pm,0})|0\rangle$ are coherent states of the two nonorthogonal modes $\hat{b}_{\pm,0}$, respectively. The $P_{\pm}(\alpha)$ are arbitrary probability distributions with $\int d\alpha P_{\pm}(\alpha)|\alpha|^2 = \epsilon$ and $N=M\epsilon$ is the total number of photons. This description applies, e.g., to incoherent superpositions of two weak thermal~\cite{Tsang} or coherent light sources.

To describe the statistics of intensity measurements in an arbitrary spatial basis we introduce the basis of detector modes $\{v_k(\bm{r})\}$ with associated field operators $\{\hat{a}_k\}$. We can express the detector modes $\hat{a}_k= \sum_{i}f_{\pm,ki}\hat{b}_{\pm,i}$ as a function of either one of the two shifted Hermite-Gauss bases with
\begin{align}\label{eq:modeoverlap}
f_{\pm,ki}=\int d^2\bm{r}v^*_k(\bm{r})u_{i}(\bm{r}\mp\bm{r}_0).
\end{align}
The average photon numbers, i.e., the expectation values of $\hat{N}_k= \hat{a}^{\dagger}_k\hat{a}_k$ can be easily determined by making use of $\langle \alpha|_{\pm}\hat{N}_k|\alpha\rangle_{\pm}=|f_{\pm,k0}|^2|\alpha|^2$ and, after measuring all $M$ copies of $\hat{\rho}(d,\theta)$, we obtain
\begin{align}\label{eq:Nk}
N_k=M\langle\hat{N}_k\rangle_{\hat{\rho}(d,\theta)}=\frac{N}{2}(|f_{+,k0}|^2+|f_{-,k0}|^2)
\end{align}
photons in mode $k$. We assume that the number of photons in each mode is measured with high precision, e.g., by using photon-counting detectors, and the parameter $d$ is estimated from this data. The precision limit of any unbiased estimator is defined by the Cram\'er-Rao bound~\cite{Kay,Lehmann,Supp}: $(\Delta d_{\mathrm{est}})\geq 1/\sqrt{N F(d,\theta)}$, where
\begin{align}\label{eq:FI}
F(d,\theta)=\sum_{k=(0,0)}^{(Q,Q)}p(k|d,\theta)\left(\frac{\partial}{\partial d}\ln p(k|d,\theta)\right)^2
\end{align}
is the Fisher information of a single photon with detection probabilities $p(k|d,\theta)=\frac{N_k}{N}$, and $Q$ is the largest index of the measured modes in both spatial dimensions. This bound can be achieved asymptotically, e.g., by a maximum likelihood estimation~\cite{Lehmann,Kay}. Through the dependence of $F(d,\theta)$ on $p(k|d,\theta)$, the achievable precision limit is determined by the measurement basis. Maximizing the Fisher information over all physically implementable measurements $\{\hat{\Pi}_k\}$ gives rise to the quantum Fisher information $F_q[\hat{\rho}(d,\theta)]=\max_{\{\hat{\Pi}_k\}} F(d,\theta)$~\cite{BC94}. For the estimation of $d$, it can be shown that $F_q[\hat{\rho}(d,\theta)]=w^{-2}$ and an ideal intensity measurement in the Hermite-Gauss basis centered at the origin, $v_k(\bm{r})=u_k(\bm{r})$, achieves this bound in the limit $Q\rightarrow\infty$ \cite{Tsang}. For small separations $d\ll 2w$, it suffices to measure the contribution of the first excited mode, $Q=1$, to saturate the quantum bound. Since the Fisher information stays finite and constant, these results imply that the precision of an estimation of $d$ is independent of the actual value of $d$, and hence, arbitrarily small distances can be resolved equally well as large ones. However, these conclusions only apply to ideal measurements without any noise and imperfections.

\begin{figure}[tb]
\centering
\includegraphics[width=.49\textwidth]{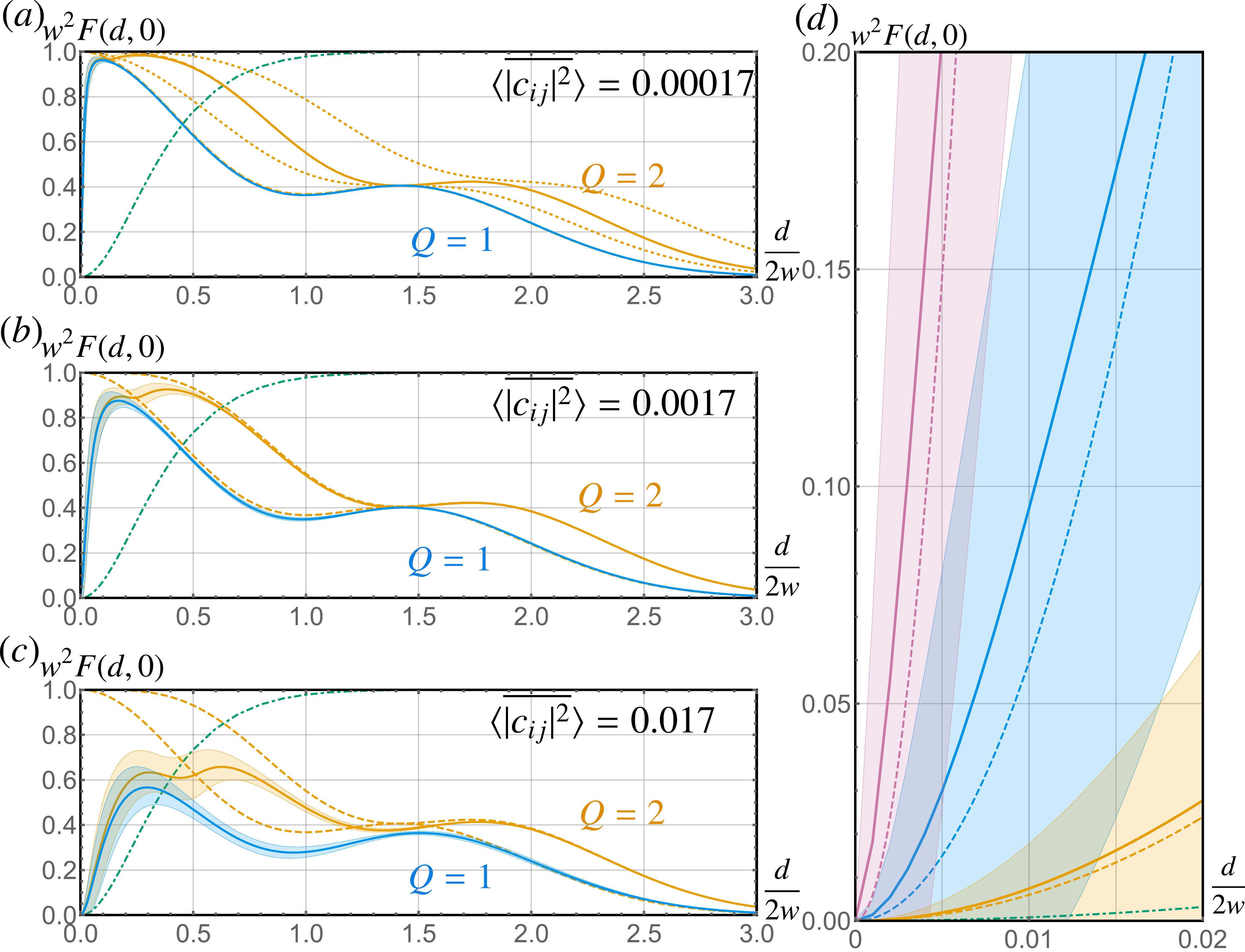}
\caption{Average Fisher information generated by a sample of 500 weak random crosstalk matrices of (a) low (average crosstalk probability $\langle \overline{|c_{ij}|^2}\rangle=0.00017$), (b) medium ($\langle \overline{|c_{ij}|^2}\rangle=0.0017$) and (c) high ($\langle \overline{|c_{ij}|^2}\rangle=0.017$) crosstalk among $D=9$ modes for measurements up to $Q=1$ (blue) and $Q=2$ (orange) modes in two dimensions. The solid lines and bands represent the average and one standard deviation. Dashed lines correspond to the ideal measurement at $\theta=0$ and the dotted lines in panel (a) show the ideal measurement at $\theta=\pi/4$. The quantum Fisher information $F_Q(d,\theta)=w^{-2}$ is reached by the ideal measurement for $Q\rightarrow\infty$ or at very small distances for $Q=1$. The green dot-dashed line describes the Fisher information for direct imaging. (d) Closeup of the breakdown of the Fisher information ($Q=1$) in the presence of nonzero crosstalk for low (orange), medium (blue), and high (violet) crosstalk. The dashed lines are the analytical predictions~(\ref{eq:FweakfixednumberC}) of the uniform crosstalk model with $|r|^2=\langle \overline{|c_{ij}|^2}\rangle$.}
\label{fig:FCross}
\end{figure}

\textit{Precision in the presence of mode crosstalk.---}To establish the impact of unavoidable deviations from the ideal mode decomposition, we determine the achievable sensitivity limits in the presence of crosstalk. We model crosstalk between the detector modes by a unitary coupling matrix $c_{kl}$ that maps the actual measurement basis to $v_{k}(\bm{r})=\sum_{l}c_{kl}u_{l}(\bm{r})$, rather than the ideal Hermite-Gauss modes $u_l(\bm{r})$; see Fig.~\ref{fig:setup}~(b). To model weak crosstalk, we consider coupling matrices whose off-diagonal elements are small compared to the diagonal ones. By including couplings into higher-order modes that are not measured, this model can effectively also describe the effect of losses. The relevant mode overlap functions~(\ref{eq:modeoverlap}) that determine the measurement statistics via Eq.~(\ref{eq:Nk}) are given by $f_{\pm,k0}=\sum_{l}c^*_{kl}\beta_{l}(\pm\bm{r}_0)$, where the $\beta_{l}(\bm{r}_0)=\int d^2\bm{r}u^*_l(\bm{r})u_{0}(\bm{r}-\bm{r}_0)$ describe the ideal scenario. The precision limits are then obtained by using the corresponding measurement data in Eq.~(\ref{eq:FI})~\cite{Supp}. 

In order to assess the impact of generic, weak crosstalk among $D$ modes, we sample randomly generated unitary crosstalk matrices $c_{ij}=C(\mu)_{ij}$ from $\mathrm{SU}(D)$ as $C(\mu)=\exp(-i\mu\sum_{k=1}^{D^2-1}\lambda_kG_k)$, where $\{G_1,\dots,G_{D^2-1}\}$ are the generalized $D\times D$ Gell-Mann matrices and the real coefficients $\lambda_1,\dots,\lambda_{D^2-1}$ are chosen randomly with normalization $\sum_{k=1}^{D^2-1}\lambda_k^2=1$~\cite{Supp}. For $\mu \ll 1$, the matrix $C(\mu)\approx \mathbb{I}-i\mu\sum_{k=1}^{D^2-1}\lambda_kG_k$ is close to the identity matrix $\mathbb{I}$ and the measurement basis is described as a small deviation from the ideal decomposition. The average crosstalk probability is determined by the average off-diagonal matrix element $\overline{|c_{ij}|^2}=\sum_{k,l=1;\:k\neq l}^D|c_{kl}|^2/D(D-1)$. Figure~\ref{fig:FCross} shows the averaged Fisher information over sets of 500 random crosstalk matrices generated with the same value of $\mu$ with $D=9$ \cite{Supp}. The ensemble-averaged crosstalk probability $\langle \overline{|c_{ij}|^2}\rangle = 0.0017$ [Fig.~\ref{fig:FCross}~(b)] corresponds to the average crosstalk experimentally measured in Ref.~\cite{Pauline}, and, for comparison, we also show the effect of crosstalk that is ten times weaker [Fig.~\ref{fig:FCross}~(a)] or stronger [Fig.~\ref{fig:FCross}~(c)]. In the limit $d/2w\rightarrow 0$, any nonzero crosstalk causes the Fisher information $F(d,\theta)$ to drop from its ideal value $w^{-2}$ to zero and then to grow approximately quadratically as $d/2w$ increases. As we will explore in detail below, this limits our ability to resolve small distances at large $N$. Yet, even in the presence of strong crosstalk, the mode decomposition achieves better sensitivities at small $d$ than direct imaging (green dot-dashed line)~\cite{Tsang,Supp}.

To analytically understand the average behavior at small separations, we introduce a uniform crosstalk model that consists of a $D\times D$ unitary matrix with entries $t$ on the diagonal and $r$ on the off diagonal, satisfying $|t|^2+(D-1)|r|^{2}=1$. For weak crosstalk probabilities $|r|^2\ll 1$ and small separations $d\ll 2w$, the Fisher information for any $Q\geq 1$ is given by~\cite{Supp}
\begin{align}\label{eq:FweakfixednumberC}
w^2F(d,\theta)&\approx\left(\frac{d}{2w}\right)^2\left(\frac{3+\cos(4\theta)}{4}\right)\frac{1}{|r|^{2}}.
\end{align}
The predictions of this model are shown in Fig.~\ref{fig:FCross}~(d) as dashed lines. Note that in contrast to crosstalk, the tilt angle $\theta$ poses no fundamental limitation to resolution since it only affects the proportionality factor $1/2\leq[3+\cos(4\theta)]/4\leq 1$.

\begin{figure}[tb]
\centering
\includegraphics[width=.46\textwidth]{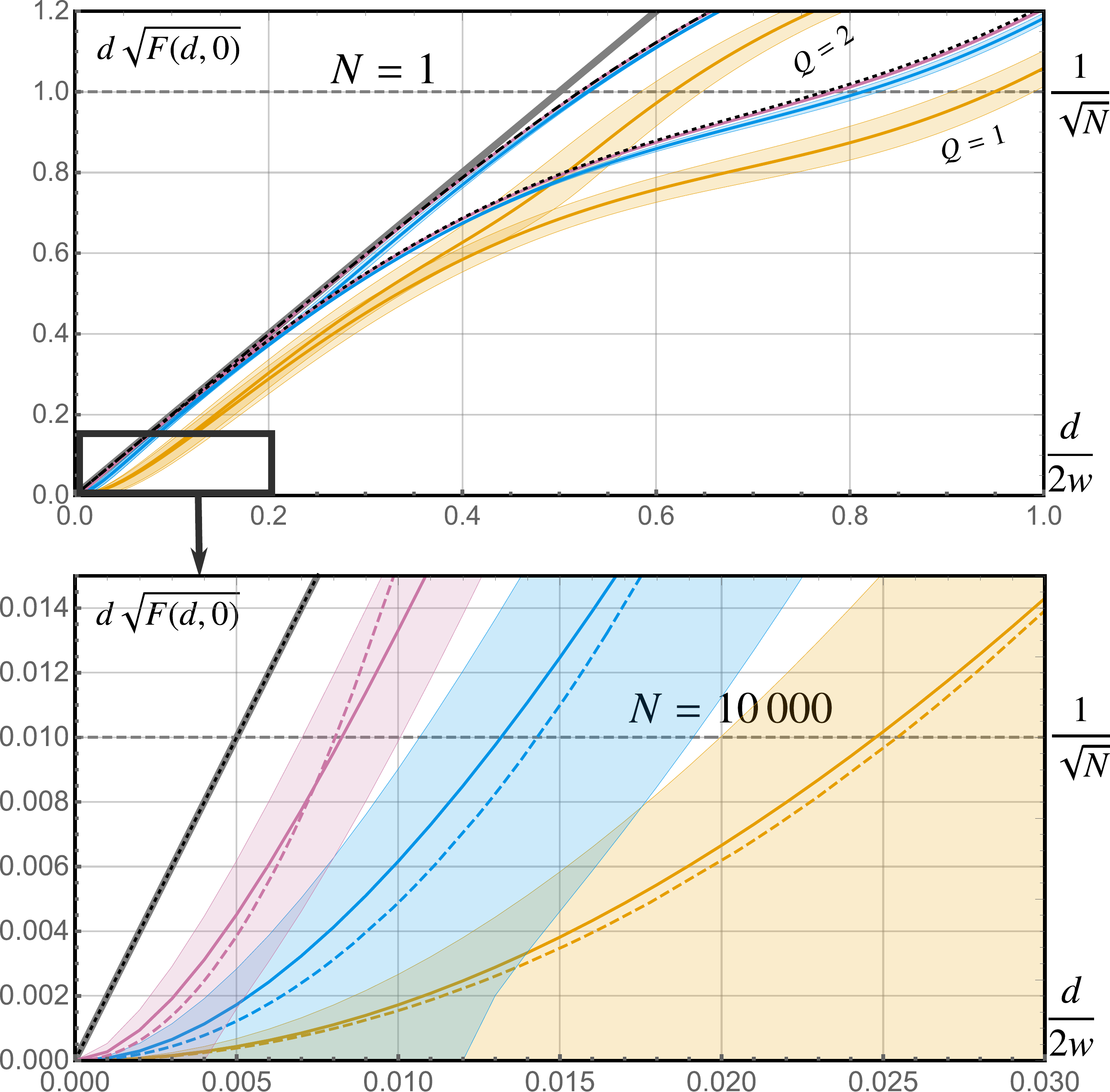}
\caption{We show $d\sqrt{F(d,0)}$ extracted from the average Fisher information (see Fig.~\ref{fig:FCross}) within one standard deviation for random uniform crosstalk with low (violet), medium (blue), and high (orange) crosstalk probability. The minimal resolvable distance $d_{\min}$ is given by the intersection of $d\sqrt{F(d,0)}$ with $1/\sqrt{N}$, where $N$ is the number of photons. In the upper panel, we show both the sensitivity for measurements up to $Q=1$ (lower lines) and $Q=2$ (upper lines). The black lines show the ideal sensitivity in the absence of crosstalk for $Q=1$ (dotted) and $Q=2$ (dot-dashed). The thick, gray line shows the ultimate quantum limit, which is achieved for $Q\rightarrow \infty$ without crosstalk. The minimal resolvable distance at $N=1$ photon is given when $d\sqrt{F(d,0)}$ intersects $1/\sqrt{N}=1$ (dashed gray line, upper panel). For small values of $d$, it suffices to restrict to measurements of $Q=1$ and we compare the sensitivity in the presence of crosstalk to the analytical prediction~(\ref{eq:FweakfixednumberC}) of the uniform crosstalk model, with $|r|^2=\langle \overline{|c_{ij}|^2}\rangle$ (dashed colored lines, lower panel). The minimal resolvable distance at $N=10\,000$ photons is found when $d\sqrt{F(d,0)}$ intersects $1/\sqrt{N}=0.01$ (dashed gray line, lower panel; cf. Figs.~\ref{fig:dminr2} and~\ref{fig:dminvsN}).}
\label{fig:dSqrtF}
\end{figure}

\begin{figure}[tb]
\centering
\includegraphics[width=.48\textwidth]{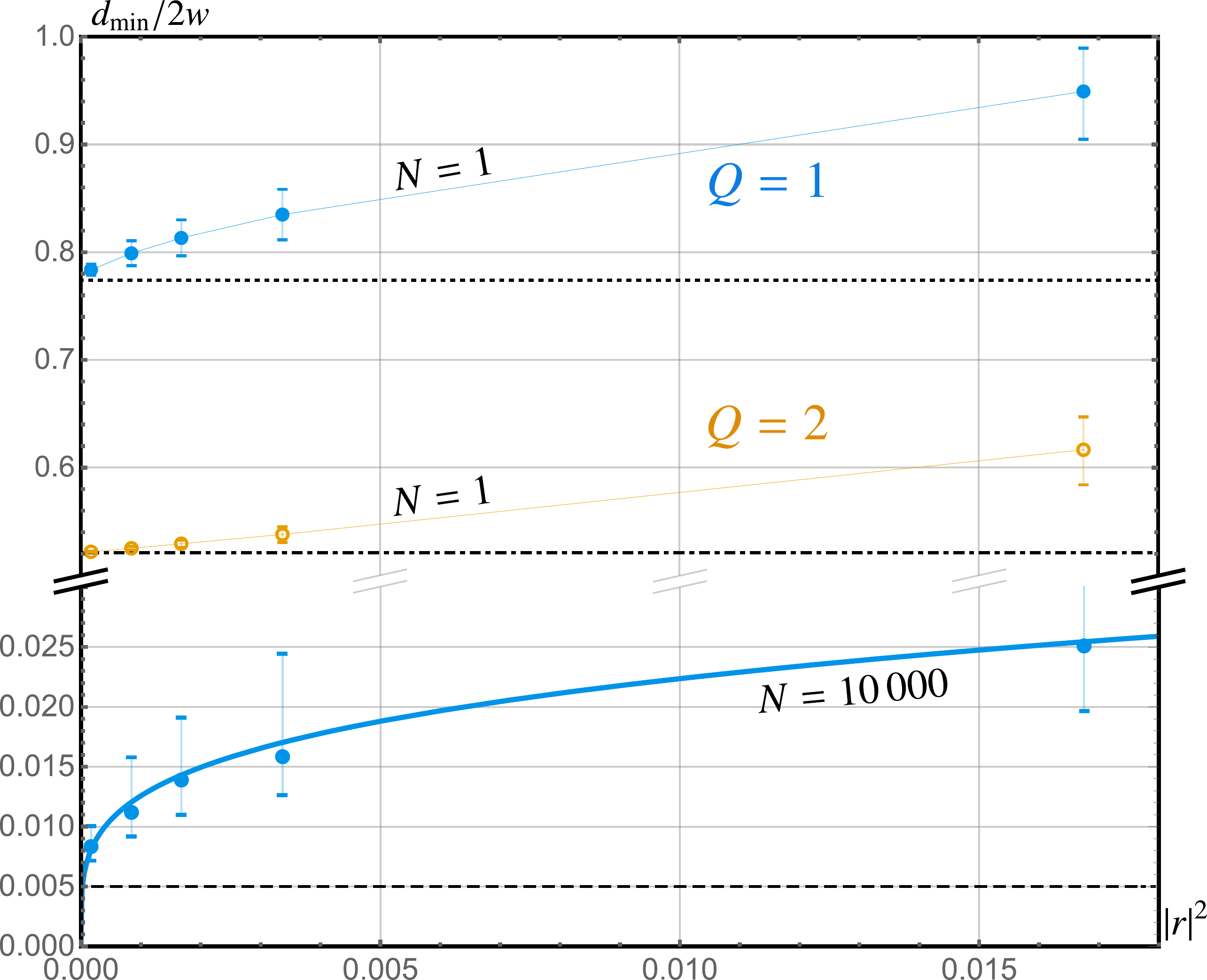}
\caption{Minimal resolvable distance $d_{\min}$ at $N=1$ (top) and $N=10\,000$ photons (bottom) as a function of the crosstalk probability $|r|^2$, obtained by intersecting $d\sqrt{F(d,0)}$ in Fig.~\ref{fig:dSqrtF} with $1/\sqrt{N}$. The dots and errors bars represent the average and standard deviation of the random crosstalk model with average off-diagonal elements $\langle \overline{|c_{ij}|^2}\rangle=|r|^2$. For $N=1$ (top) we show the results for measurements up to $Q=1$ (blue) and $Q=2$ (orange). The black horizontal lines show $d_{\min}$ for a measurement without crosstalk for $N=1$ with $Q=1$ (dotted) and $Q=2$ (dot-dashed), and the ideal quantum limit~(\ref{eq:dminideal}), $Q\rightarrow\infty$, yields $d_{\min}/2w=0.5$ in this case. At $N=10\,000$ (bottom) there is hardly any improvement by measuring $Q=2$ or higher, and we only show $Q=1$. The thick blue line shows the analytical prediction~(\ref{eq:dminP}) of the uniform crosstalk model at $\theta=0$. A crosstalk-free measurement with $N=10\,000$ and $Q=1$ yields a $d_{\min}$ that cannot be distinguished from the quantum limit $d_{\min}/2w=0.005$ (dashed) on this scale.}
\label{fig:dminr2}
\end{figure}

\textit{Minimal resolvable distance.---}The minimal distance between emitters that can still be resolved is determined by the signal-to-noise ratio (SNR) and requires that $\mathrm{SNR}(d)=d/(\Delta d)\geq 1$. An efficient, unbiased estimator~\cite{Lehmann} minimizes the noise by saturating the Cram\'er-Rao bound, leading to $\Delta d=1/\sqrt{NF(d,\theta)}$, and we obtain
\begin{align}\label{eq:SNR}
\mathrm{SNR}(d)=d\sqrt{NF(d,\theta)}.
\end{align}
We thus define the minimal resolvable distance as the smallest solution $d_{\min}$ to $\mathrm{SNR}(d_{\min})=1$. In Fig.~\ref{fig:dSqrtF}, we show $d\sqrt{F(d,\theta)}$ as a function of $d$ for weak, random crosstalk. For a given number $N$ of photons, the minimal resolvable distance is identified as the intersection with $1/\sqrt{N}$. For large photon numbers, $N\gg 1$, $d_{\min}$ is dominated by the behavior of $F(d,\theta)$ in the limit of $d/2w\rightarrow 0$. In the case of an ideal measurement (considering either $Q\rightarrow\infty$ or $d\ll 2w$ with any $Q\geq 1$), the Fisher information is constant, $F(d,\theta)= w^{-2}$~\cite{Tsang}. Hence, in the absence of crosstalk, we obtain the characteristic ``shot-noise'' scaling
\begin{align}\label{eq:dminideal}
d_{\min}=\frac{w}{\sqrt{N}}.
\end{align}
In contrast, the quadratic dependence of $F(d,\theta)$ on $d$ observed in Eq.~(\ref{eq:FweakfixednumberC}) modifies the scaling with the total number of photons $N$ and we obtain in the presence of crosstalk,
\begin{align}\label{eq:dminP}
d_{\min}=\frac{w}{N^{\frac{1}{4}}}\sqrt{2|r|}\left(\frac{4}{3+\cos(4\theta)}\right)^{\frac{1}{4}}.
\end{align}
The minimal resolvable distance for $N=10\,000$ photons is shown in the lower part of Fig.~\ref{fig:dminr2}. The scaling of the average predictions of the random crosstalk model agrees with that of the analytical, uniform crosstalk model~(\ref{eq:dminP}) at the same average crosstalk probability.

\begin{figure}[t!]
\centering
\includegraphics[width=.48\textwidth]{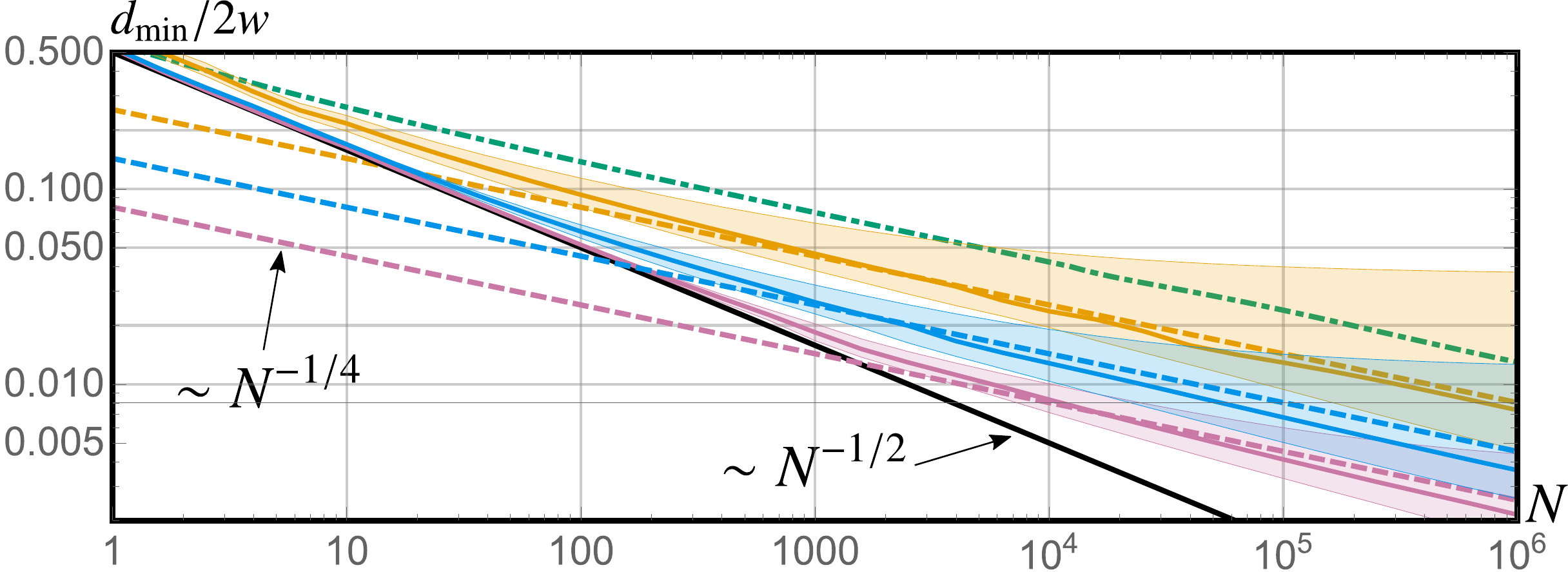}
\caption{Scaling of the minimal resolvable distance $d_{\min}$ with the number $N$ of photons in the presence of random uniform crosstalk with low (violet), medium (blue) and high (orange) crosstalk probability (cf. Fig.~\ref{fig:FCross}) for measurements up to $Q=2$. For small $N\lesssim 10$, the minimal resolvable distance $d_{\min}$ still follows the ideal $N^{-1/2}$ scaling of Eq.~(\ref{eq:dminideal}) (black line), with a prefactor that increases with the crosstalk. For larger $N$, the scaling changes and approaches $N^{-1/4}$, as predicted analytically in Eq.~(\ref{eq:dminP}) by the uniform crosstalk model and is plotted with $|r|^2=\langle \overline{|c_{ij}|^2}\rangle$ (dashed lines). The green dot-dashed line shows the scaling of $d_{\min}$ for an ideal direct imaging measurement.}
\label{fig:dminvsN}
\end{figure}

At large $N$, the quadratic scaling of $d\sqrt{F(d,\theta)}$ (see the lower panel in Fig.~\ref{fig:dSqrtF}) leads to the $N^{-1/4}$-scaling in Eq.~(\ref{eq:dminP}). In contrast, at small $N\approx 1$, we find that $d_{\min}$ is determined by the behavior of $F(d,\theta)$ at finite values of $d/2w$ (see the upper panel of Fig.~\ref{fig:dSqrtF}). In this case, the leading term in an expansion of $F(d,\theta)= c+\mathcal{O}(d)$ is independent of $d$, and thus $d\sqrt{F(d,\theta)}$ scales approximately linearly, which implies $d_{\min}\approx 1/\sqrt{Nc}$, i.e., the $N^{-1/2}$-scaling observed in Eq.~(\ref{eq:dminideal}). The same is true at arbitrary $N$ for the estimation of small deviations $d$ from a fixed separation $d_0\gg d$: In this case, $F(d+d_0,\theta)$ is constant to leading order in $d$ and yields an $N^{-1/2}$ scaling for $d_{\min}$.

As these cases are dominated by the behavior of $F(d,\theta)$ at values of $d\approx 2w$, the measurement of higher excited modes becomes increasingly important and can yield significant advantages also in the presence of crosstalk, which no longer poses a fundamental limitation on the precision (see upper panel of Fig.~\ref{fig:dSqrtF}). The minimal resolvable distance with $N=1$ photon is shown in the upper part of Fig.~\ref{fig:dminr2} as a function of the crosstalk. A measurement of $Q=2$ comes close to the ideal quantum resolution limit~(\ref{eq:dminideal}) even in the presence of crosstalk.

In the presence of crosstalk, the scaling of $d_{\min}$ (see Fig.~\ref{fig:dminvsN}) changes from the ideal $N^{-1/2}$ dependence [Eq.~(\ref{eq:dminideal})] at small values of $N$ to a much less favorable $N^{-1/4}$ scaling [Eq.~(\ref{eq:dminP})] in the experimentally relevant regime of large $N$. This behavior is confirmed by the statistical data of the random crosstalk model. For typical crosstalk probabilities, the transition occurs already at moderate photon numbers of $N\approx 10^1-10^4$. The resolvable distance for direct imaging measurements shows an analogous change of scaling~\cite{Supp} but it is outperformed even by very noisy mode decompositions.

\textit{Conclusions.}---We have identified the precision limits for an estimation of the separation of two incoherent point sources by intensity measurements after a realistic spatial mode decomposition. Introducing a general model for measurement crosstalk and losses using basis transformations, we have compared statistical data from random crosstalk matrices to analytical results obtained from a uniform model with tunable crosstalk probability. We observe that within statistical error margins, the scaling of the precision limits with the number of photons is device independent as it only depends on the average crosstalk probability. The uniform crosstalk model further allows us to analytically derive these scaling laws.

To quantify the precision for this estimation, we considered the smallest separation $d_{\min}$ that can be distinguished from the noise. This definition naturally depends on the number of measured photons. The most relevant information for practical measurements with many photons is contained in the large-$N$ scaling. Crosstalk, however small, will always be present in realistic experimental mode decompositions and leads to a significant change of scaling from $N^{-1/2}$ to $\alpha N^{-1/4}$ with a prefactor $\alpha$ that depends on the average crosstalk probability. In contrast, measurements at low photon numbers $N\leq 10$ are less affected by crosstalk, and in this case $d_{\min}$ increases by a crosstalk-dependent factor without changing the $N^{-1/2}$ scaling.

\textit{Acknowledgments.}---We would like to thank K. Banaszek, P. Boucher, C. Datta, and Y.-L. Len for discussions. M.G. acknowledges funding by the LabEx ENS-ICFP: ANR-10-LABX-0010/ANR-10-IDEX-0001-02 PSL*. N.T. acknowledges financial support of the Institut Universitaire de France. This work was supported the European Union's Horizon 2020 research and innovation programme under the QuantERA programme through the project ApresSF.

\clearpage
\begin{center}
{\large \bfseries Supplementary Material}
\end{center}

\setcounter{equation}{0} \setcounter{figure}{0} \renewcommand\thefigure{S%
\arabic{figure}} \renewcommand\theequation{S\arabic{equation}}

In Section~\ref{app:FIideal}, we derive the Fisher information for the estimation of $d$ from an ideal measurement of up to $Q$ Hermite-Gauss modes for arbitrary separations $d$ and tilt angles $\theta$. In Section~\ref{app:directimaging}, we recall the Fisher information for direct imaging. In Section~\ref{app:modelscross}, we introduce models for crosstalk and in Section~\ref{app:cross}, we derive the Fisher information including the effect of crosstalk. Finally, in Section~\ref{sec:MPLC}, we briefly describe how spatial mode decompositions may be implemented.

\section{Fisher information: Ideal measurement}\label{app:FIideal}
\subsection{Fisher information for Poissonian distributed intensity measurements in a finite set of modes}
We begin by briefly recalling the derivation of the Fisher information for a Poissonian photodetection model, see, e.g., Ref.~\cite{Chao}. Assume that intensity measurements of a finite set of detector modes $1,\dots,K$ are performed in order to estimate the parameter of interest. Photons in modes of higher order $k>K$ are not detected. This gives rise to events of the kind $x=(n_1,\dots,n_K)$, where $n_1,\dots,n_K$ denote the number of registered clicks in each of the respective modes. We further assume no correlations or bunching effects between the recorded photons, such that the statistics in each mode $k$ is given by a Poissonian distribution with average $N_k$, respectively. This yields the following probability for the event $x$:
\begin{align}\label{eq:multipoisson}
p(n_1,\dots,n_K|d,\theta)=\frac{1}{n_1!\cdots n_K!}N_1^{n_1}\cdots N_K^{n_K}e^{-N_D},
\end{align}
where $N_D=\sum_{k=1}^KN_k$ is the total number of detected photons. Each of the average values $N_k$ depends on the parameters $d$ and $\theta$. With ideal detectors we recover the total number of photons as $N=\lim_{K\rightarrow\infty}N_D$.

The Fisher information for estimations of the parameter $d$ with this measurement is given by
\begin{align}\label{eq:PoiFIStart}
\mathcal{F}(d,\theta)=\left\langle \left(\frac{\partial}{\partial d}\log p(n_1,\dots,n_K|d,\theta)\right)^2\right\rangle
\end{align}
where the average of an arbitrary function $f(n_1,\dots,n_K)$ is obtained from~(\ref{eq:multipoisson}) as
\begin{align}
\left\langle f(n_1,\dots,n_K)\right\rangle&=\sum_{n_1=0}^{\infty}\cdots\sum_{n_K=0}^{\infty}p(n_1,\dots,n_K|d,\theta) f(n_1,\dots,n_K).\notag
\end{align}
To determine~(\ref{eq:PoiFIStart}), we use~(\ref{eq:multipoisson}) to write
\begin{align}
\log p(n_1,\dots,n_K|d,\theta)=\sum_{k=1}^K\left(\log(-n_k!)+n_k\log N_k-N_k\right),\notag
\end{align}
and we obtain
\begin{align}\label{eq:dddlog}
\frac{\partial}{\partial d}\log p(n_1,\dots,n_K|d,\theta)&=\sum_{k=1}^K\left(n_k\frac{\partial}{\partial d}\log N_k-\frac{\partial}{\partial d}N_k\right)\notag\\
&=\sum_{k=1}^K\left(\frac{n_k}{N_k}-1\right)\frac{\partial}{\partial d} N_k.
\end{align}
Inserting~(\ref{eq:dddlog}) into Eq.~(\ref{eq:PoiFIStart}), we find
\begin{align}
\mathcal{F}(d,\theta)
&=\sum_{k,l=1}^K\left(\frac{\langle n_kn_l\rangle}{N_kN_l}-\frac{\langle n_k\rangle}{N_k}-\frac{\langle n_l\rangle}{N_l}+1\right)\left(\frac{\partial}{\partial d} N_k\right)\left(\frac{\partial}{\partial d} N_l\right).\notag
\end{align}
Making use of the Poissonian average and variance,
\begin{align}
\langle n_k\rangle &= N_k,\\
\langle n_kn_l\rangle &= N_kN_l + \delta_{kl}N_k,
\end{align}
we obtain
\begin{align}\label{eq:fullF}
\mathcal{F}(d,\theta)&=\sum_{k=1}^K\frac{1}{N_k}\left(\frac{\partial}{\partial d} N_k\right)^2.
\end{align}
We define
\begin{align}\label{eq:pkdN}
p(k|d,\theta)=N_k/N
\end{align}
as the probability for a single photon to be detected in mode $k$, such that~(\ref{eq:fullF}) reads 
\begin{align}\label{eq:Fisheradditive}
\mathcal{F}(d,\theta)=NF(d,\theta),
\end{align}
where
\begin{align}\label{eq:Fdtheta}
F(d,\theta)&=\sum_{k=1}^K\frac{1}{p(k|d,\theta)}\left(\frac{\partial}{\partial d} p(k|d,\theta)\right)^2
\end{align}
is the Fisher information associated with a single photon in the form of Eq.~(4) in the main text. Notice that the probability obeys the normalization 
\begin{align}
\sum_{k=1}^{\infty}p(k|d,\theta)=1,
\end{align}
but only the first $K$ terms contribute to Eq.~(\ref{eq:Fdtheta}). Since each term in the sum~(\ref{eq:Fdtheta}) is non-negative, we see that the Fisher information increases as more modes are measured.

In practical situations, rather than the total number $N$ of emitted photons, only the number $N_D$ of detected photons may be known. In contrast to $N$, this number may in principle depend on the parameter $d$. Defining $p_D(k|d,\theta)=N_k/N_D$ and following the same approach as above yields from Eq.~(\ref{eq:fullF}):
\begin{align}
\mathcal{F}(d,\theta)&=N_DF_D(d,\theta)+(2N_D+1)\frac{1}{N_D}\left(\frac{\partial}{\partial d} N_D\right)^2,
\end{align}
where $F_D(d,\theta)=\sum_{k=1}^K\frac{1}{p_D(k|d,\theta)}\left(\frac{\partial}{\partial d} p_D(k|d,\theta)\right)^2$.
Since the second term is non-negative, the bound $\mathcal{F}(d,\theta)\geq N_DF_D(d,\theta)$ holds. This implies that we may effectively replace $N$ by $N_D$ in Eqs.~(\ref{eq:pkdN}) and~(\ref{eq:Fdtheta}): With this replacement, Eq.~(\ref{eq:Fisheradditive}) generally becomes a lower bound on the actual Fisher information, and this bound is tight when $N_D$ is independent of $d$. For measurements in the Hermite-Gauss basis and small values of $d$ this is a good approximation since only few photons go by undetected in highly excited modes with $k>K$.

\subsection{Mode overlap and statistics}
We consider the ideal measurement in a basis of Hermite-Gauss modes, $u_k(\bm{r})=u_{nm}(\bm{r})$ with $k=(n,m)$. In this case, we can write the overlap integrals as $f_{\pm,k0}=\beta_{nm}(\pm\bm{r}_0)$, where
\begin{align}\label{eq:overlapideal}
\beta_{nm}(\bm{a}):=\int d^2\bm{r}u^*_{nm}(\bm{r})u_{00}(\bm{r}-\bm{a}).
\end{align}
The Hermite-Gauss function are defined for $\bm{r}=(x,y)$ as
\begin{align}
u_{nm}(x,y)=\frac{1}{\sqrt{(\pi/2)w^22^{n+m}n!m!}}H_n\left(\sqrt{2}\frac{x}{w}\right)H_m\left(\sqrt{2}\frac{y}{w}\right)e^{-\frac{x^2+y^2}{w^2}},
\end{align}
where $H_n(x)$ are the Hermite polynomials and $w$ is the point spread function that determines the width of the $00$-mode as
\begin{align}
u_{00}(x,y)=\frac{\sqrt{2}}{w\sqrt{\pi}}e^{-\frac{x^2+y^2}{w^2}}.
\end{align}
The integral~(\ref{eq:overlapideal}) corresponds to the overlap of a coherent state of a two-dimensional quantum harmonic oscillator (displaced to phase space coordinates $\bm{a}$) with the excited state that contains $n$ and $m$ excitations in the respective directions. We use polar coordinates to express $x$ and $y$ in function of their separation $d$ and tilt angle $\theta$. Using polar coordinates has the advantage that $d$ represents the distance between emitters for arbitrary values of $\theta$, whereas in cartesian coordinates, a nonzero tilt requires the estimation of the nonlinear function $\sqrt{x^2+y^2}$ of the parameters $x$ and $y$. In polar coordinates, we obtain
\begin{align}\label{eq:betanm}
\beta_{nm}(\bm{r}_0)=\frac{1}{\sqrt{n!m!}}\left(\frac{d}{2w}\right)^{n+m}(\cos\theta)^n(\sin\theta)^me^{-\frac{1}{2}\left(\frac{d}{2w}\right)^2}.
\end{align}
We obtain the average photon numbers in each mode as
\begin{align}
N_{nm}=\frac{N}{2}\left(|\beta_{nm}(\bm{r}_0)|^2+|\beta_{nm}(-\bm{r}_0)|^2\right)=N|\beta_{nm}(\bm{r}_0)|^2.
\end{align}
Moreover, the probability to find a detector click in the mode $nm$ is given by
\begin{align}\label{eq:probIdeal}
p(nm|d,\theta)&=\frac{N_{nm}}{N}=|\beta_{nm}(\bm{r}_0)|^2.
\end{align}
This probability is conditioned on the true value of the displacement being $\bm{r}_0$. In this scenario, the problem of resolving two incoherent sources is equivalent to the estimation of the position of a single emitter.

\subsection{Fisher information for the estimation of the emitter distance}
We first make use of Eq.~(\ref{eq:betanm}) to obtain
\begin{align}
\frac{\partial}{\partial d}\beta_{nm}(\pm\bm{r}_0)&=\frac{1}{d}\left(n+m-\left(\frac{d}{2w}\right)^2\right)\beta_{nm}(\pm\bm{r}_0).
\end{align}
Since $\beta_{nm}(\bm{r})\in\mathbb{R}$, this implies that
\begin{align}
\frac{\partial}{\partial d}p(nm|d,\theta)&=2\left(\frac{\partial}{\partial d}\beta_{nm}(\bm{r}_0)\right)\beta_{nm}(\bm{r}_0)\notag\\
&=\frac{2}{d}\left(n+m-\left(\frac{d}{2w}\right)^2\right)\beta_{nm}(\bm{r}_0)^2.
\end{align}
Assuming that all modes $nm$ with $0\leq n\leq Q$ and $0\leq m\leq Q$ are measured, we obtain the Fisher information
\begin{align}
F(d,\theta)&=\sum_{n,m=0}^Q\frac{1}{p(nm|d,\theta)}\left(\frac{\partial}{\partial d}p(nm|d,\theta)\right)^2\label{eq:Fdgeneral}\\
&=\sum_{n,m=0}^Q\frac{4}{d^2}\left(n+m-\left(\frac{d}{2w}\right)^2\right)^2\beta_{nm}(\bm{r}_0)^2.
\end{align}
Substituting
\begin{align}
x=\frac{d}{2w}
\end{align}
and inserting Eq.~(\ref{eq:betanm}) leads to
\begin{align}\label{eq:Fd}
&\quad F(d,\theta)\\
&=\sum_{n,m=0}^Q\frac{1}{n!m!}\frac{x^{2(n+m-1)}}{w^2}\left(n+m-x^2\right)^2(\cos\theta)^{2n}(\sin\theta)^{2m}e^{-x^2}\notag.
\end{align}
We obtain the limits
\begin{align}
&\lim_{Q\rightarrow\infty} F_{d}(d,\theta)=\frac{1}{w^2}
\end{align}
and for any $Q>0$:
\begin{align}\label{eq:smalldistancelimit}
&\lim_{d\rightarrow0} F_{d}(d,\theta)=\frac{1}{w^2},
\end{align}
which are independent of $\theta$ and correspond to the quantum Fisher information $F_Q[d,\theta]=w^{-2}$~\cite{Tsang}. 

\begin{figure}[tb]
\centering
\includegraphics[width=.48\textwidth]{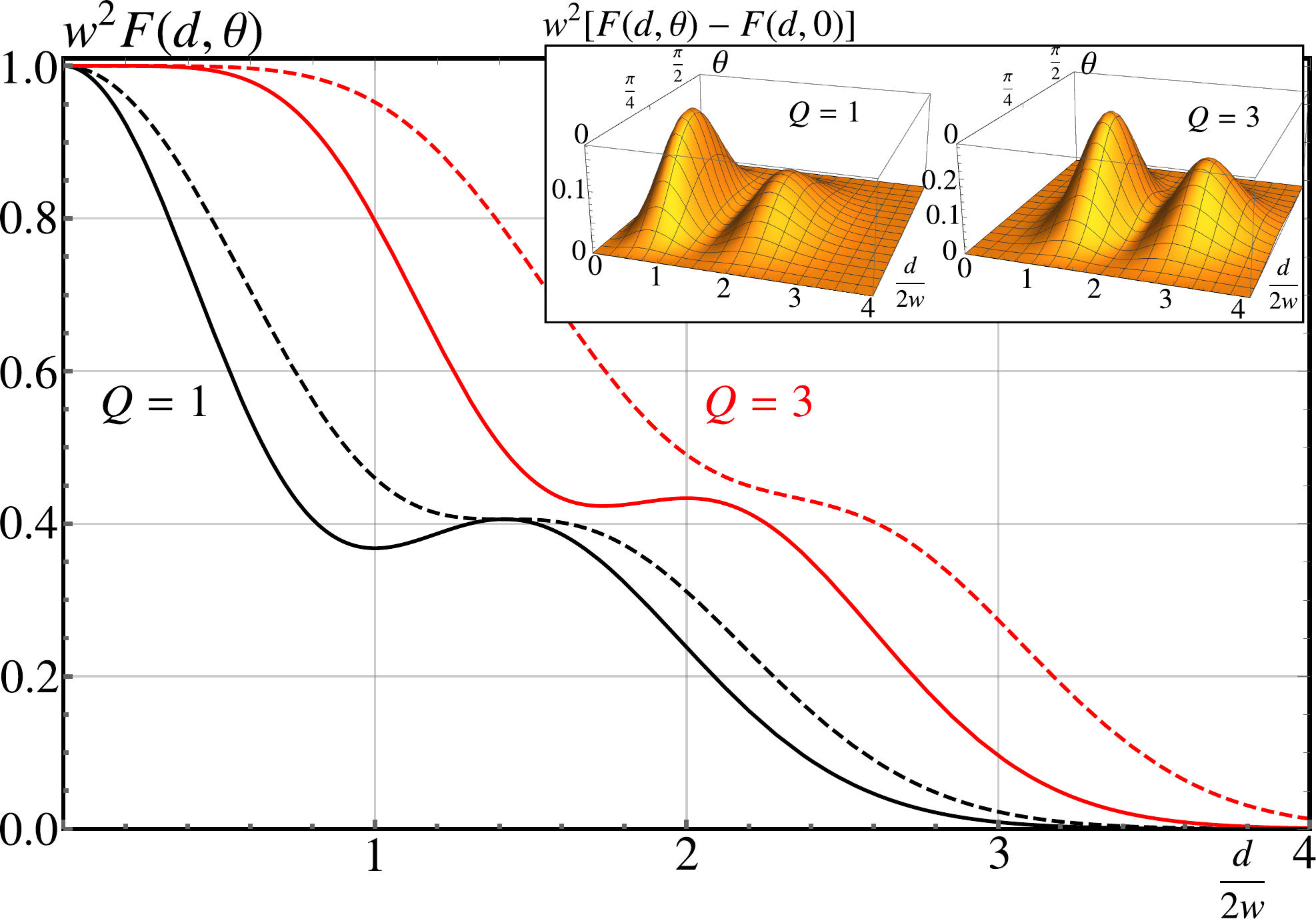}
\caption{Fisher information $w^2 F(d,\theta)$ for the estimation of the separation $d$ in a two-dimensional setup with a tilt angle $\theta$ between the separation axis of the two sources and the measurement apparatus as a function of $d/2w$. We show the sensitivity $F(d,\theta)$ at its lowest ($\theta=0$ or $\theta=\pi/2$, continuous lines) and highest ($\theta=\pi/4$, dashed lines) values, with $Q=1$ (black) and $Q=3$ (red). The inset shows the difference $F(d,\theta)-F(d,0)$ as a function of $d/2w$ and $\theta$.}
\label{fig:align}
\end{figure}

Moreover, we find the following explicit expressions for specific values of the orientation angle $\theta$:
\begin{align}\label{eq:resultsFd0}
F(d,0)&:=\lim_{\theta\rightarrow0} F(d,\theta)=\lim_{\theta\rightarrow\frac{\pi}{2}} F(d,\theta)\\&=\frac{1}{w^2Q!}\left(e^{-x^2}(x^2-(Q+1))x^{2Q}+\Gamma(Q+1,x^2)\right),\notag
\end{align}
where $\Gamma(Q+1,x)$ is the upper incomplete Gamma function, which, for positive integer $Q$ satisfies the inequality~\cite{Arfken2012}
\begin{align}
\Gamma(Q+1,x):=\int_x^{\infty} e^{-t}t^Qdt=Q!e^{-x}\sum_{k=0}^Q\frac{x^k}{k!}.
\end{align}
As we can see from Fig.~\ref{fig:align}, in these limits we achieve the lowest sensitivity over all angles $\theta$, while the maximal sensitivity is achieved at $\theta=\pi/4$.

\section{Direct imaging}\label{app:directimaging}
An ideal direct imaging measurement estimates the source separation from the intensity distribution in the image plane. The average intensity at position $\bm{r}$ is
\begin{align}
I(\bm{r})&=M\langle \hat{E}^{(+)}(\bm{r})^{\dagger}\hat{E}^{(+)}(\bm{r})\rangle\notag\\
&=\frac{M}{2}\int d\alpha P_+(\alpha)\langle\alpha|_{+} \hat{E}^{(+)}(\bm{r})^{\dagger}\hat{E}^{(+)}(\bm{r})|\alpha\rangle_{+}\notag\\
&\quad + \frac{M}{2}\int d\alpha P_-(\alpha)\langle\alpha|_{-} \hat{E}^{(+)}(\bm{r})^{\dagger}\hat{E}^{(+)}(\bm{r})|\alpha\rangle_{-}\notag\\
&=\frac{N}{2}\left(|u_{00}(\bm{r}-\bm{r}_0)|^2+|u_{00}(\bm{r}+\bm{r}_0)|^2\right),
\end{align}
where we used $\hat{E}^{(+)}(\bm{r})|\alpha\rangle_{\pm}=\alpha u_{00}(\bm{r}\mp\bm{r}_0)|\alpha\rangle_{\pm}$. Using the Poissonian distribution of photons, this leads to the Fisher information~\cite{Tsang}
\begin{align}
\mathcal{F}_{\rm{DI}}(d,\theta)&=\int d\bm{r}\frac{1}{I(\bm{r})}\left(\frac{\partial}{\partial d}I(\bm{r})\right)^2,
\end{align}
or, equivalently, $\mathcal{F}_{\rm{DI}}(d,\theta)=NF_{\rm{DI}}(d,\theta)$, with
\begin{align}
F_{\rm{DI}}(d,\theta)&=\int d\bm{r}\frac{1}{p(\bm{r}|d,\theta)}\left(\frac{\partial}{\partial d}p(\bm{r}|d,\theta)\right)^2,
\end{align}
where $p(\bm{r}|d,\theta)=I(\bm{r})/N$. This integral can be evaluated numerically and is independent of $\theta$ (see green dot-dashed lines in Fig.~2 in the main manuscript and Fig.~\ref{fig:FCrossSupp} below). For $x=d/2w\ll 1$, we can approximate it analytically as
\begin{align}\label{eq:DIFIsmallx}
w^2F_{\rm{DI}}(d,\theta)=8x^2+\mathcal{O}(x^4).
\end{align}

\section{Models for crosstalk}\label{app:modelscross}
\subsection{Random crosstalk}
\begin{figure}[tb]
\centering
\includegraphics[width=.48\textwidth]{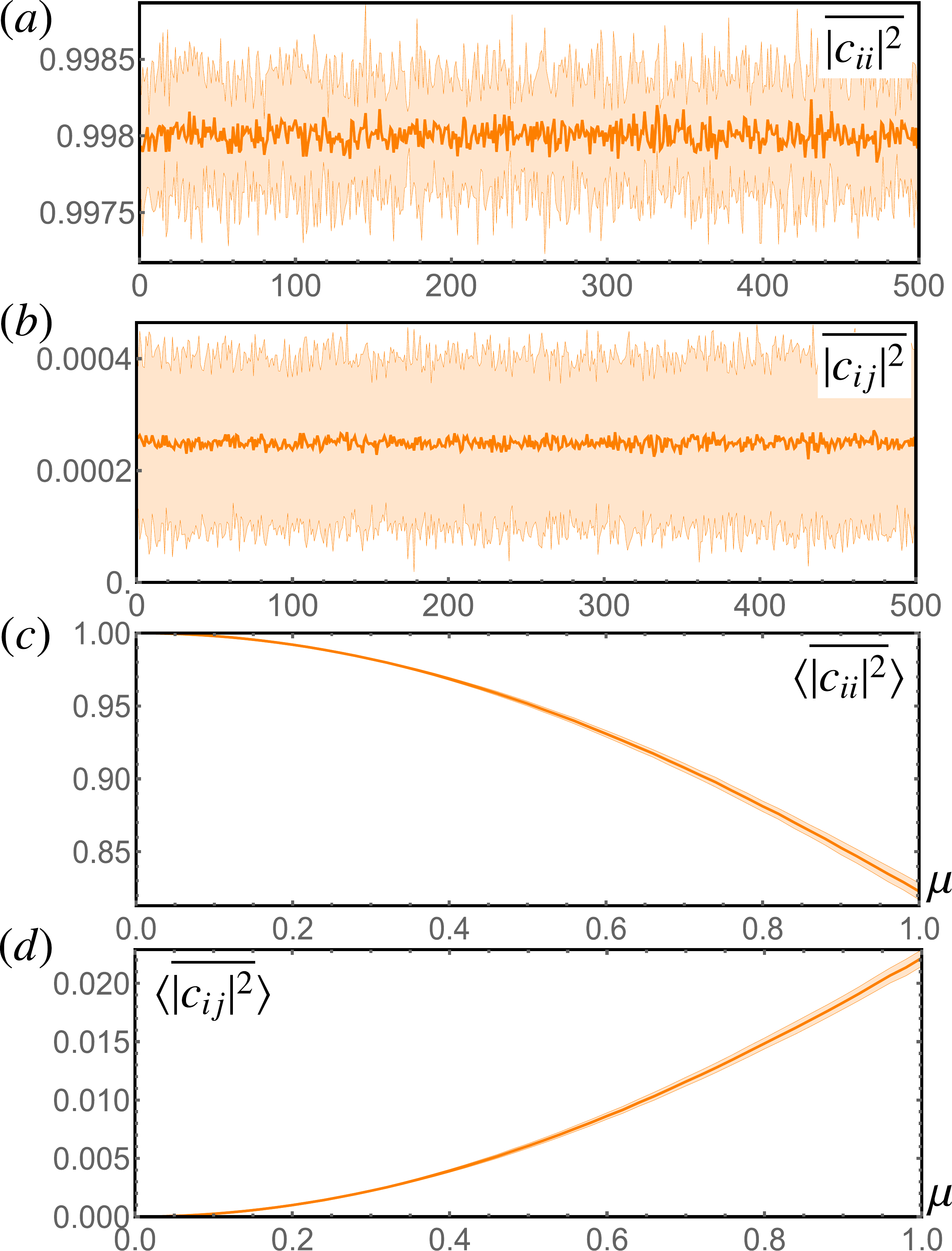}
\caption{Statistics of weak random crosstalk matrices. The distribution of the average absolute squared diagonal (a) and off-diagonal (b) elements including one standard deviation of a sequence of 500 randomly sampled $9\times 9$ matrices $C(\mu)$ with $\mu=0.1$ shows that matrices sampled with the same parameter have very similar crosstalk probability. Panels (c) and (d) show the same properties, averaged over 500 matrices as a function of $\mu$.}
\label{fig:benchmarkRndU}
\end{figure}
We introduce a random-matrix model for unitary crosstalk matrices from a basis $\{G_1,\dots,G_{D^2-1}\}$ of the Lie algebra $\mathfrak{su}(D)$, the generalized $D\times D$ Gell-Mann matrices~\cite{GellMann2008}. We generate a random unitary $D\times D$ matrix by sampling the random, real coefficients $\lambda_1,\dots,\lambda_{D^2-1}$ with $\sum_{k=1}^{D^2-1}\lambda_k^2=1$ producing a unitary matrix 
\begin{align}
C(\mu)=\exp(-i\mu\sum_{k=1}^{D^2-1}\lambda_kG_k),
\end{align}
with fixed $\mu>0$. 

We characterize these random matrices, $C(\mu)_{ij}=c_{ij}$, by analyzing the average absolute square of diagonal and off-diagonal elements, respectively: For each random matrix, we define
\begin{align}
\overline{|c_{ii}|^2}=\frac{1}{D}\sum_{k=1}^D|c_{kk}|^2\\
\overline{|c_{ij}|^2}=\frac{1}{D(D-1)}\sum_{\substack{k,l=1\\k\neq l}}^D|c_{kl}|^2.
\end{align}
Unitary ensures that
\begin{align}
\overline{|c_{ii}|^2}+(D-1)\overline{|c_{ij}|^2}=1.
\end{align}
Figure~\ref{fig:benchmarkRndU}~(a) and~(b) shows $\overline{|c_{ii}|^2}$ and $\overline{|c_{ij}|^2}$ for 500 random matrices with fixed value $\mu$. These quantities have relatively low fluctuations around a well-defined average value which is determined by $\mu$. Figure~\ref{fig:benchmarkRndU}~(c) and~(d) displays the dependence of the ensemble averaged values $\langle\overline{|c_{ii}|^2}\rangle$ and $\langle\overline{|c_{ij}|^2}\rangle$ over 500 random matrices on the parameter $\mu$. The quantity $\langle\overline{|c_{ij}|^2}\rangle$ represents the average probability for crosstalk into a specific mode.

\subsection{Uniform crosstalk}
A simple analytical crosstalk model is given by the $D\times D$ uniform coupling matrix
\begin{align}\label{eq:uniformC}
C=\begin{pmatrix} t & r & \dots & r\\
r &t & & r\\
\vdots &&\ddots & \vdots\\
r & \dots & r & t
\end{pmatrix},
\end{align}
with $|t|^2+(D-1)|r|^2=1$ and
\begin{align}\label{eq:pscat}
P_{\rm{scat}}=1-|t|^2=(D-1)|r|^2
\end{align}
is the overall probability to scatter into an undesired mode. The limit of weak uniform crosstalk is described by $|r|^2\ll 1$ at fixed $P_{\rm{scat}}$. When the additional condition $|r|^2 \ll 1/(D-1)$ is satisfied (as is the case in the simulations shown in the main text), the weak crosstalk limit also implies that $P_{\rm{scat}}\ll 1$. 

\subsection{Generic crosstalk}
The uniform model can be generalized to a generic crosstalk model by considering a coupling matrix with arbitrary entries $c_{ij}$. The weak crosstalk limit corresponds to the limit $|c_{ij}|\ll 1$ for all $i\neq j$ at constant $|c_{ii}|$. The only assumption in this model is that the off-diagonal elements are small compared to the diagonal ones.

\section{Fisher information: Presence of crosstalk}\label{app:cross}
Crosstalk is modelled by a unitary coupling matrix that describes the actual measurement basis as a linear combination of the ideal basis. Expressed with two-dimensional indices, $k=(n,m)$, we obtain
\begin{align}
v_{nm}(\bm{r})=\sum_{kl}c_{nm,kl}u_{kl}(\bm{r}),
\end{align}
with ideal Hermite-Gauss modes $u_{kl}(\bm{r})$. We obtain the average photon numbers
\begin{align}
N_{nm}&=\frac{N}{2}\left(|\gamma_{nm}(\bm{r}_0)|^2+|\gamma_{nm}(-\bm{r}_0)|^2\right),
\end{align}
where $\gamma_{nm}(\bm{r})=\sum_{kl}c_{nm,kl}^*\beta_{kl}(\bm{r})$ and $\beta_{kl}(\bm{r})$ are the ideal overlap functions defined in Eq.~(\ref{eq:overlapideal}). Using again $p(nm|d,\theta)=N_{nm}/N$, we find the probability distribution
\begin{align}\label{eq:probcross}
&\quad p(nm|d,\theta)\notag\\&=\frac{1}{2}\sum_{klpq}c^*_{nm,kl}c_{nm,pq}\left(\beta_{kl}(\bm{r}_0)\beta_{pq}(\bm{r}_0)+\beta_{kl}(-\bm{r}_0)\beta_{pq}(-\bm{r}_0)\right)\notag\\
&=\sum_{klpq}c^*_{nm,kl}c_{nm,pq}\frac{1+(-1)^{k+p+l+q}}{2}\beta_{kl}(\bm{r}_0)\beta_{pq}(\bm{r}_0)\notag\\
&=\sum_{\substack{klpq\\k+l+p+q\in 2 \mathbb{N}}}c^*_{nm,kl}c_{nm,pq}\beta_{kl}(\bm{r}_0)\beta_{pq}(\bm{r}_0),
\end{align}
where we used that $\beta_{nm}(\bm{r})\in\mathbb{R}$. Notice that due to
\begin{align}
\beta_{kl}(-\bm{r}_0)\beta_{pq}(-\bm{r}_0)
&=(-1)^{k+l+p+q}\beta_{kl}(\bm{r}_0)\beta_{pq}(\bm{r}_0),
\end{align}
only terms where $k+l+p+q\in 2 \mathbb{N}$ is an even number contribute. We obtain the derivative
\begin{align}\label{eq:pdercrossXtheta}
&\quad\frac{\partial}{\partial d}p(nm|d,\theta)
\\&=\sum_{\substack{klpq\\k+l+p+q\in 2 \mathbb{N}}}c^*_{nm,kl}c_{nm,pq}\frac{1}{d}\left(k+l+p+q-2\left(\frac{d}{2w}\right)^2\right)\notag\\&\hspace{1.5cm}\times\beta_{kl}(\bm{r}_0)\beta_{pq}(\bm{r}_0),\notag
\end{align}
which permits us to calculate the Fisher information for arbitrary crosstalk matrices using Eq.~(\ref{eq:Fdgeneral}).

\subsection{Approximation for small displacements in the presence of crosstalk}
To study the limitations imposed by crosstalk on the Fisher information at small source separation, we perform a perturbative expansion of $F(d,\theta)$, assuming
\begin{align}
x=\frac{d}{2w}\ll 1.
\end{align}
We obtain from Eq.~(\ref{eq:betanm})
\begin{align}
\beta_{00}(\bm{r}_0)&=1-\frac{1}{2}x^2+\mathcal{O}(x^4)\notag\\
\beta_{10}(\bm{r}_0)&= x\cos\theta+\mathcal{O}(x^3)\notag\\
\beta_{01}(\bm{r}_0)&= x\sin\theta+\mathcal{O}(x^3)\notag\\
\beta_{11}(\bm{r}_0)&=x^2(\cos\theta)(\sin\theta)+\mathcal{O}(x^4)\notag\\
\beta_{nm}(\bm{r}_0)&=\mathcal{O}(x^3)\quad\forall n+m\geq 3.
\end{align}
According to Eq.~(\ref{eq:probcross}) we find
\begin{align}
p(nm|d,\theta)&=|c_{nm,00}|^2+\mathcal{O}(x^2).
\end{align}
Moreover, from Eq.~(\ref{eq:pdercrossXtheta}) follows
\begin{align}
&\quad w\frac{\partial}{\partial d}p(nm|d,\theta)\\&=\sum_{\substack{klpq\\k+l+p+q\in 2 \mathbb{N}}}c^*_{nm,kl}c_{nm,pq}\frac{1}{2x}\left(k+l+p+q-2x^2\right)\notag\\&\hspace{2cm}\times\beta_{kl}(\bm{r}_0)\beta_{pq}(\bm{r}_0),\notag\\
&=x \left[(\cos\theta)^2|c_{nm,10}|^2+(\sin\theta)^2|c_{nm,01}|^2-|c_{nm,00}|^2\right.\notag\\&\hspace{0.7cm}+\left.\sin(2\theta)\mathrm{Re}(c^*_{nm,10}c_{nm,01}+c^*_{nm,11}c_{nm,00})
\right]+\mathcal{O}(x^2).
\notag
\end{align}

\subsection{Uniform crosstalk}
For the uniform crosstalk matrix given in Eq.~(\ref{eq:uniformC}), we obtain for $Q\geq 1$:
\begin{align}
w\frac{\partial}{\partial d}p(00|d,\theta)&= x\left[|r|^2-|t|^2+g(r,t,\theta)\right]+\mathcal{O}(x^2),\notag\\
w\frac{\partial}{\partial d}p(10|d,\theta)&= x\left[(\cos\theta)^2(|t|^2-|r|^2)+g(r,t,\theta)\right]+\mathcal{O}(x^2),\notag\\
w\frac{\partial}{\partial d}p(01|d,\theta)&= x\left[(\sin\theta)^2(|t|^2-|r|^2)+g(r,t,\theta)\right]+\mathcal{O}(x^2),\notag\\
w\frac{\partial}{\partial d}p(11|d,\theta)&= xg(r,t,\theta)+\mathcal{O}(x^2),\notag\\
w\frac{\partial}{\partial d}p(nm|d,\theta)&=2 x\sin(2\theta)|r|^2+\mathcal{O}(x^2)\quad\forall n+m\geq3,\notag
\end{align}
with $g(r,t,\theta)=\sin(2\theta)(|r|^2+\mathrm{Re}(t^*r))$. The Fisher information now reads
\begin{align}\label{eq:Fdcross}
w^2F(d,\theta)&=\sum_{n,m=0}^Q\frac{1}{p(nm|d,\theta)}\left(w\frac{\partial}{\partial d}p(nm|d,\theta)\right)^2\\
&=x^2\frac{\left[|r|^2-|t|^2+g(r,t,\theta)\right]^2}{|t|^2}\notag\\
&\quad+x^2\frac{\left[(\cos\theta)^2(|t|^2-|r|^2)+g(r,t,\theta)\right]^2}{|r|^2}\notag\\
&\quad+x^2\frac{\left[(\sin\theta)^2(|t|^2-|r|^2)+g(r,t,\theta)\right]^2}{|r|^2}\notag\\
&\quad+x^2\frac{g(r,t,\theta)^2}{|r|^2}\notag\\&\quad+4x^2(Q-1)^2\sin(2\theta)^2|r|^2+\mathcal{O}(x^4).\notag
\end{align}
Simple expressions can be obtained, e.g., in the limit $\theta\rightarrow 0$ or $\theta\rightarrow \frac{\pi}{2}$, i.e., when the separation axis between the two emitters is aligned with the measurement basis. In this case, we obtain $F(d,0)=\lim_{\theta\rightarrow 0}F(d,\theta)=\lim_{\theta\rightarrow \frac{\pi}{2}}F(d,\theta)$ with
\begin{align}
w^2F(d,0)&=x^2\left(\frac{|r|^4}{|t|^2} + \frac{|t|^4}{|r|^2} - |r|^2 - |t|^2\right)+\mathcal{O}(x^4).
\end{align}
To obtain the limit of weak uniform crosstalk, we assume $|r|^2\ll 1$ at fixed $P_{\rm{scat}}=(D-1)|r|^2$, see Eq.~(\ref{eq:pscat}). We obtain
\begin{align}\label{eq:Fcrosssmall}
w^2F(d,\theta)&=x^2\left(\frac{3+\cos(4\theta)}{4}\frac{(1-P_{\rm{scat}})^2}{|r|^{2}}+\mathcal{O}(|r|^{-1})\right)+ \mathcal{O}(x^4),
\end{align}
where $\frac{1}{2}\leq\frac{3+\cos(4\theta)}{4}\leq 1$. For $P_{\rm{scat}}\ll 1$ we obtain the expression
\begin{align}\label{eq:Fcrosssmallsimpler}
w^2F(d,\theta)&=x^2\left(\frac{3+\cos(4\theta)}{4}\frac{1}{|r|^{2}}+\mathcal{O}(|r|^{-1})\right)+ \mathcal{O}(x^4),
\end{align}
that was stated in the main text.

\subsection{Generic crosstalk}
The result~(\ref{eq:Fcrosssmall}) can be generalized to generic crosstalk, i.e., we can drop the assumption that the elements of the crosstalk matrix are identical on and off the diagonal, respectively. It suffices to assume that all off-diagonal elements are small compared to the diagonal elements, i.e., the elements $c_{nm,kl}$ for all $n\neq k$ and $m\neq l$ are of the order $\epsilon$, where $0<\epsilon \ll 1$. An analogous derivation as above yields
\begin{align}\label{eq:FcrosssmallGeneric}
&\quad w^2F(d,\theta)\\
&=x^2\left(\frac{|c_{01,01}| ^4 (\sin \theta)^4}{|c_{01,00}|^2}+\frac{|c_{10,10}|^4 (\cos \theta)^4}{|c_{10,00}|^2}+\mathcal{O}(\epsilon^{-1})\right)+\mathcal{O}(x^4).\notag
\end{align}
The behavior at small $x$ is dominated by the probability to correctly transmit the first excited modes $01$ and $10$ (given by $|c_{01,01}|^2$ and $|c_{10,10}|^2$, respectively) and the probability to scatter from the first excited modes $01$ and $10$ into the mode $00$ (given by $|c_{01,00}|^2$ and $|c_{10,00}|^2$, respectively). The weight of the two directions depends on the orientation of the two sources via the angle $\theta$. We recover Eq.~(\ref{eq:Fcrosssmall}) in the special case where $|c_{01,01}|^2=|c_{10,10}|^2=|t|^2=1-P_{\rm{scat}}$ and $|c_{01,00}|^2=|c_{10,00}|^2=|r|^2$ using $(\sin \theta)^4+(\cos\theta)^4=\frac{3+\cos(4\theta)}{4}$.

\subsection{Crosstalk vs. losses}
\begin{figure}[tb]
\centering
\includegraphics[width=.49\textwidth]{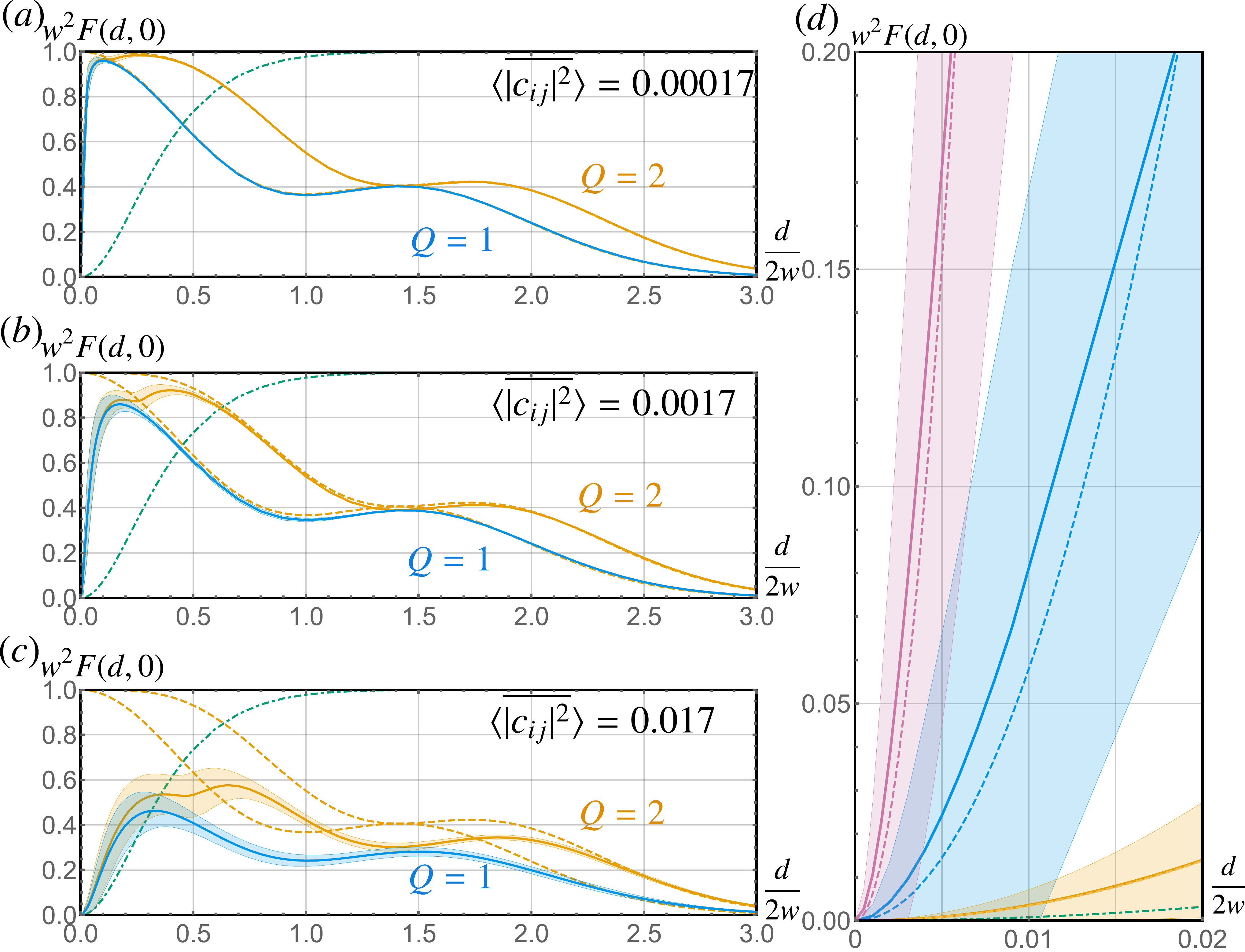}
\caption{Same as Fig.~2 in the main text, but with larger crosstalk matrices, $D=16$. The dashed lines in panel (d) are the analytical predictions of Eq.~(\ref{eq:Fcrosssmall}). The orange dashed line lies on top of the orange continuous line.}
\label{fig:FCrossSupp}
\end{figure}

The data shown in the main text describes weak uniform crosstalk with average crosstalk probabilities $\langle \overline{|c_{ij}|^2}\rangle$ of the order of $1\%$ or less at fixed $D=9$. This describes the crosstalk between the modes of a measurement with $Q=2$, i.e., $u_{nm}$ with $n,m\in\{0,1,2\}$. We compare to the analytical predictions of the uniform crosstalk model by identifying $\langle \overline{|c_{ij}|^2}\rangle=|r|^2$ and $P_{\rm{scat}}$ is typically small enough to justify Eq.~(\ref{eq:Fcrosssmallsimpler}) as approximation of Eq.~(\ref{eq:Fcrosssmall}). As long as this approximation is valid, the effect of crosstalk is independent of $D$ and depends only on the value $|r|^2$. This is confirmed also for the random crosstalk model by additional numerical data shown in Fig.~\ref{fig:FCrossSupp}, which shows the effect of random crosstalk with $D=16$ but with the same values for $\langle \overline{|c_{ij}|^2}\rangle$ (to be compared to Fig.~2 in the main text). We observe that also the overall shape of the function $F(d,\theta)$ beyond the small-$d$ approximation remains largely invariant under an increase of $D$. Deviations are visible only for high crosstalk at $\langle \overline{|c_{ij}|^2}\rangle=0.017$, which according to~(\ref{eq:pscat}) already implies a scattering probability of $P_{\rm{scat}}\approx 0.25$.

For small $d$, we compare to the prediction of Eq.~(\ref{eq:Fcrosssmall}), which contains a correction factor due to finite losses and is shown as dashed lines in Fig.~\ref{fig:FCrossSupp}~(d). The difference between~(\ref{eq:Fcrosssmall}) and~(\ref{eq:Fcrosssmallsimpler}) is less than $5\%$ for small and medium crosstalk, but becomes more significant for large crosstalk where $(1-P_{\rm{scat}})^2\approx 43\%$. With this factor, the prediction of Eq.~(\ref{eq:Fcrosssmall}) is visually indistinguishable from average random crosstalk prediction (orange line).

\subsection{Minimal resolvable distance}
The minimal resolvable distance for large $N$ is obtained from the definition in the main text and yields for the uniform crosstalk model:
\begin{align}
d^{\rm{U}}_{\min}=\frac{w}{N^{\frac{1}{4}}}\sqrt{\frac{2|r|}{1-P_{\rm{scat}}}}\left(\frac{4}{3+\cos(4\theta)}\right)^{\frac{1}{4}}.
\end{align}
For an ideal direct imaging measurement, we obtain from Eq.~(\ref{eq:DIFIsmallx}):
\begin{align}\label{eq:dminDItheo}
d^{\rm{DI}}_{\min}=\frac{w}{N^{\frac{1}{4}}}\left(\frac{1}{2}\right)^{\frac{1}{4}}.
\end{align}
We have $d^{\rm{U}}_{\min}<d^{\rm{DI}}_{\min}$, i.e., an imperfect mode decomposition outperforms ideal direct imaging as long as
\begin{align}
\frac{|r|^2}{(1-P_{\rm{scat}})^2}<\frac{1}{8}.
\end{align}
In the uniform model this corresponds to crosstalk probabilities as large as $|r|^2\approx 0.03-0.05$ with $D=9$ and $D=16$, respectively, which is roughly 20-30 times stronger than the values observed in experimental characterizations of the spatial mode sorter~\cite{Pauline}.

We can compare the scaling of $d_{\min}$ with $N$ for the two measurement strategies in Fig.~5 of the main text, where the numerically determined $d_{\min}$ for direct imaging with arbitrary $N$ is shown as the green dot-dashed line. The analytical prediction for large $N$ is given by Eq.~(\ref{eq:dminDItheo}). The numerical data follows this prediction closely for $N\gtrsim 100$. Since the two curves would be hard to distinguish visually, the analytical result is not shown in the plot.

\section{Implementing spatial mode decompositions}\label{sec:MPLC}
Spatial mode decompositions can be realized using the multi-plane-light-conversion (MPLC) technique~\cite{Labroille}. The basic idea of this technique is that any unitary basis transformation can in principle be implemented with a finite number of phase plates and spatial Fourier transforms. In the MPLC system used in Ref.~\cite{Pauline}, a decomposition into the Hermite-Gauss modes $u_{nm}$ with $n,m\in\{0,1,2\}$ is realized by sending the light through a series of carefully designed phase plates, separated by mirrors. The path length of the propagation between two consecutive phase plates is chosen such the corresponding basis change is to a good approximation described by a Fourier transform.

\end{document}